\documentclass[conference]{IEEEtran}

\IEEEoverridecommandlockouts
\usepackage{cite}
\usepackage{amsmath,amssymb,amsfonts}
\usepackage{algorithmic}
\usepackage{graphicx}
\usepackage{textcomp}
\usepackage{xcolor}
\usepackage{verbatim}
\def\BibTeX{{\rm B\kern-.05em{\sc i\kern-.025em b}\kern-.08em
    T\kern-.1667em\lower.7ex\hbox{E}\kern-.125emX}}
\begin{document}

\title{A Methodology for Controlling the Emotional Expressiveness in Synthetic Speech - a Deep Learning approach \\
}

\author{\IEEEauthorblockN{No\'e Tits}
\IEEEauthorblockA{\textit{Numediart Institute} \\
\textit{University of Mons}\\
Mons, Belgium \\
noe.tits@umons.ac.be}

}

\maketitle

\begin{abstract}
In this project, we aim to build a Text-to-Speech system able to produce speech with a controllable emotional expressiveness. We propose a methodology for solving this problem in three main steps. The first is the collection of emotional speech data. We discuss the various formats of existing datasets and their usability in speech generation. The second step is the development of a system to automatically annotate data with emotion/expressiveness features. We compare several techniques using transfer learning to extract such a representation through other tasks and propose a method to visualize and interpret the correlation between vocal and emotional features. The third step is the development of a deep learning-based system taking text and emotion/expressiveness as input and producing speech as output. We study the impact of fine tuning from a neutral TTS towards an emotional TTS in terms of intelligibility and perception of the emotion.


\end{abstract}










\section{Introduction \& Motivations}

Text-To-Speech (TTS) synthesis systems, which synthesize speech from text, have been around for decades and have improved very recently with the advent of new machine learning techniques such as Deep Neural Networks (DNN). Companies such as Google and Amazon provide free DNN-based speech synthesis systems such as WaveNet\footnote{https://cloud.google.com/text-to-speech/}~\cite{wavenet-16-vandenoord} or Amazon Polly\footnote{https://aws.amazon.com/fr/polly/}. These systems offer an excellent quality of speech, with voices obtained by analyzing tens of hours of neutral speech, therefore not very expressive.

The challenge faced by researchers today has therefore evolved towards the the field of expressive speech synthesis~\cite{emotional_speech_synthesis-14-Burkhardt}. 

The objective of this project is to produce, not a neutral voice, but remarkable voices, similar to those of the actors, possessing a specific grain and a great capacity for expressiveness. This will make it possible to create virtual agents communicating in a more natural way, and thus to improve the quality of the interaction with a machine, by making it closer to a human-human interaction.

In what follows, the background and related work of the field is briefly summarized and remaining challenges are emphasized. 

\label{related_work}

Depending on the synthesis technique used~\cite{emotional_speech_synthesis-14-Burkhardt}, the voice is more or less natural and the synthesis parameters are more or less numerous. These parameters allow to create variations in the voice. The number of parameters is therefore important for the synthesis of expressive speech.

While formant synthesis can control many parameters, the resulting voice is unnatural. Synthesizers using the principle of concatenation of speech segments seem more natural but allow the control of few parameters.

The statistical approaches based on the use of HMM allow to obtain a natural synthesis as well as a control of many parameters~\cite{statistical_param_speech_syn-09-zen}. The most recent statistical approach uses DNN~\cite{stat_param_speech_synthesis_dnn-13-zen} and is the basis of new speech synthesis systems such as WaveNet~\cite{wavenet-16-vandenoord} and Tacotron~\cite{tacotron-17-wang}. The improvement provided by this technique~\cite{hmms_to_dnns-16-watts} comes from the replacement of decision trees by DNNs and the replacement of state prediction (HMM) by frame prediction. This improvement is very interesting for the expression of emotions.


There has been numerous studies to assess DNN based systems ability to synthesize speech with high quality and naturalness such as Wavenet~\cite{wavenet-16-vandenoord}, Tacotron~\cite{tacotron-17-wang},  Deep Convolutional TTS~\cite{dctts-17-tachibana} (DCTTS), WaveRNN~\cite{wavernn-18-Kalchbrenner}, Char2Wav~\cite{char2wav-17-sotelo} and Deep Voice~\cite{deep-voice-17-arik}. However, important issue in the literature is the need of tens of hours of speech data and a lot of computational power. 

DCTTS needs less computational resources than the others. In~\cite{dctts-17-tachibana}, they have suggested that they were able to train a TTS model in 15 hours using a regular computer with two GPUs, resulting in nearly acceptable speech synthesis.

A closer look to the literature on control of speech quality and emotional content, however, reveals a number of gaps. Access to a large amount of data remains a problem. The recording of high quality emotional speech datasets is expensive and time consuming. The amount of data available is therefore relatively limited compared to the needs of deep learning algorithms. This problem has previously been addressed using promising methods related to knowledge transfer such as transfer learning~\cite{transfer_learning-2010-pan}, fine-tuning and multi-task learning.


Concerning the controllable aspect of TTS system, an important issue is the labelization of speech data with style or emotional information.
Recent studies were conducted on techniques of unsupervised learning to achieve controllable speech synthesis without the need of labels.

In~\cite{tacotron_prosody-18-skerry}, the authors investigated on an extension to the Tacotron speech synthesis system. The technique they propose is a system that learns a latent embedding space by encoding audio features into a vector that is concatenated to text information and then fed to Tacotron. These latent embeddings model the remaining variability of speech after accounting for variation due to phonetics, speaker identity, and channel effects.

In~\cite{expressive_speech_vae-18-akuzawa}, a Variational Auto-encoder (VAE) is combined with VoiceLoop, a TTS system.

A series of recent studies have used the concept of VAE~\cite{generative_controllable_speech-18-hsu,unsupervised_controllable_speech-18-henter,style_tokens-18-wang}. In~\cite{generative_controllable_speech-18-hsu}, they 
combine a VAE with Gaussian Mixture Model (GMM) and name it GMVAE. In~\cite{unsupervised_controllable_speech-18-henter}, they combine Vector Quantization with a VAE (VQ-VAE).

These works has provided evidence that building a latent space can lead to variables useful to control style in speech synthesis.
In~\cite{generative_controllable_speech-18-hsu}, they demonstrate that their system can synthesize spectrograms with different rhythms, speaking rates and $F_0$'s from a single text, proving a control on these parameters.

A weakness of these studies is that do not provide insights about the relationships between the resulting latent space and the audio features possible to control. We aim to fill this gap.

Section~\ref{plan} exposes the methodology we propose to achieve the objectives of the project in three main points. 
The contributions made to these three tasks are presented in Section~\ref{datasets},~\ref{emotion} and~\ref{synthesis}. 
Section~\ref{datasets} addresses the problem of data collection. 
Section~\ref{emotion} presents techniques of emotion representation and their applicability for the control of speech synthesis.
Section~\ref{synthesis} considers the problem of speech synthesis with emotional content.
Finally, Section~\ref{ccl} concludes and  
describes the remaining challenges and plans for the future of this project.

\section{Plan}
\label{plan}

To synthesize expressive speech, the main problem is the variability of the vocal expression of emotions. Deep Learning has been effective at handling complex data but requires a large amount of these annotated data. The difficulty is then to annotate large databases with expressive metadata that is very subjective and not well defined.

The goal is to create an automatic system for annotating large expressive vocal databases, and synthesizing expressive speech. 

Our work plan contains 3 main tasks:
\begin{itemize}
    \item Collect emotional speech data
    \item Build a system able to extract a representation of emotional expressiveness in speech
    \item Build a system able to synthesize expressive speech based on the data collected and the representation extracted
\end{itemize}



This mapping will be used to create a tool that will control the expressiveness of synthesized speech by varying the dimensions continuously.
This system will then be integrated into a speech synthesis instrument from which one will be able to play an utterance as he wishes.


In a second part of the thesis, we will focus on the automatic control of the expressive dimensions. The goal will then be to synthesize expressive speech based on the text to be produced and the emotion of the interlocutor. The tool must respond to the interlocutor imitating its expressiveness.



\section{Datasets}
\label{datasets}

As of today, there are a lot of speech datasets available online. A good source of speech data are audiobooks (LJ-Speech\footnote{https://keithito.com/LJ-Speech-Dataset/}, LibriSpeech~\cite{librispeech-15-panayotov}, LibriTTS~\cite{libritts-19-zen}) in which we have automatically transcriptions corresponding to speech. 

Other datasets come from research projects. Some are designed for speech generation purpose (voice conversion~\cite{vctk-17-veaux}, speech synthesis~\cite{cmuArctic04speechDB, siwis17speechDB, AmuS17emoDB}) while others are designed for a speech analysis goal (ASR, emotion in speech etc.). The last often consist of conversational setups and contain overlaps in speech as well as noise compared to datasets for speech generation. This is the reason why they cannot be used for emotional speech synthesis.

These datasets contain a fair amount of data for the use of deep learning algorithms. However, either they are not suited for synthesis purpose, or they are poor in expressiveness.

Other datasets offer emotionally rich content with a high quality, but in a limited amount~\cite{ravdess18emoDB, cremad14emoDB, GEMEP12emoDB, berlinEmo05emoDB}.

To fill this gap, we collected a speech dataset called EmoV-DB Database~\cite{emov-db-18-tits} rich in expressiveness and suited for speech generation purpose in a fair amount to use deep learning algorithms.

It contains data for male and female actors in English and a male actor in French. 
The English actors were recorded in two different anechoic chambers of the Northeastern University campus while the French actor was recorded in an anechoic chamber of the University of Mons.
Each actor was asked to utter a subset of the sentences from the CMU-Arctic database for English speakers and from the SIWIS database for the French speaker. 
The database covers 5 classes (amusement, anger, disgust, sleepiness and neutral) with the goal to build, in the future, synthesis and voice transformation systems with the potential to control the emotional dimension in a continuous way.

It is available online and there is a script available for using gentle's forced alignment~\footnote{https://github.com/numediart/EmoV-DB}.

\begin{table}[htbp]
\begin{center}
\caption{Description of EmoV-DB}
\resizebox{\columnwidth}{!}{
\begin{tabular}{|p{0.3\columnwidth}|p{0.7\columnwidth}|}
    \hline
    Type of data & Audio, text and emotion category\\ \hline
    How data was acquired & Audio recorded in anechoic chambers of the Northeastern University campus. \\ \hline
    Data format & Segmented in sentences, associated with transcriptions (CMU-Artic/SIWIS), classified in emotional categories\\ \hline
    Experimental features & Recordings of sentences uttered by 2 male and 2 female speakers in 5 different emotions, making a total of 7000 sentences\\ \hline
    Data accessibility & https://github.com/numediart/EmoV-DB \\
    \hline
    \end{tabular}
}
\end{center}
\end{table}

\section{Emotion representation}
\label{emotion}

Despite the great improvements that signal processing field has known, the field of affective computing remains challenging. As mentioned in Section~\ref{plan}, we want to build a system able to extract a representation of emotional expressiveness in speech. In this Section we present two studies. The first one~\cite{asr-based-features-18-tits} investigates if a DNN trained on an Automatic Speech Recognition (ASR) task learns features useful for predicting emotion dimensions (valence, arousal). 

The second one~\cite{visualization-19-tits} investigates a good way to represent expressiveness in speech without emotional label. We evaluate these representations with a correlation analysis with the eGeMAPS feature set~\cite{egemaps-16-eyben}, a feature set designed for affective computing tasks.

\subsection{ASR-based features for Emotion recognition}

In this experiment, we study whether a deep learning based ASR can be used as a feature extractor for an emotion recognition task. An ASR system is a mapping between audio features and text features. To have a good accuracy, an ASR has to take into account the variability in speech to be able to recover text information from all kinds of speaker identity, style etc. Therefore, an intermediate representation extracted by the system may contain useful information to predict emotion in speech.

We used an open source ASR available online~\footnote{https://github.com/buriburisuri/speech-to-text-wavenet}. It consists of a stack of 15 dilated convolutional layers. Each layer is constituted by 128 gated convolutional units (GCU) with skip connection. This architecture is inspired by Wavenet~\cite{wavenet-16-vandenoord}. The system was trained with MFCCs extracted from VCTK dataset~\cite{vctk-17-veaux}.

The analyzed data comes from IEMOCAP database~\cite{iemocap-08-busso}. It contains 5 sessions of dialogues between a male and a female, making a total of 10 speakers. These conversations were acted either with a script or improvised from a given situation described to the actors. The dataset is segmented in utterances and each utterance was annotated in emotional dimensions (valence, arousal, dominance) and emotion categories.

The analysis reveals that predicting valence and arousal with a linear model trained based on the output of GCUs outperform a similar model based on the eGeMAPS~\cite{egemaps-16-eyben} features extracted with opensmile~\footnote{https://www.audeering.com/opensmile/}, the state of the art of handcrafted features for emotion recognition.

\subsection{Interpretation of latent spaces}

In this Section, the aim is to investigate ways to represent expressiveness in speech without the need of any labels that are difficult and expensive to collect. Moreover, it is of interest to have an interpretable representation to be able to build controllable speech synthesis systems with an understandable and predictable behaviour.

We present a methodology to obtain latent spaces, visualize them and analyze their relationship with audio features by correlation. 
We then explain a way to visualize the influence of variation of embeddings on audio features.

We compare three techniques of encoding mel-spectrogram. The principle is to encode the mel-spectrogram by training deep learning algorithms on three different tasks. The encoding is used for:
\begin{itemize}
    \item classification in categories of style,
    \item speaker classification,
    \item conditioning a Text-to-Speech synthesis along with the text.
\end{itemize}

The mel-spectrogram is encoded to vector of length of 8 that contains expressiveness information.
This vector is computed for each utterance of the dataset to obtain a collection of embeddings.

\begin{figure}[h]
  \includegraphics[width=1.07\linewidth]{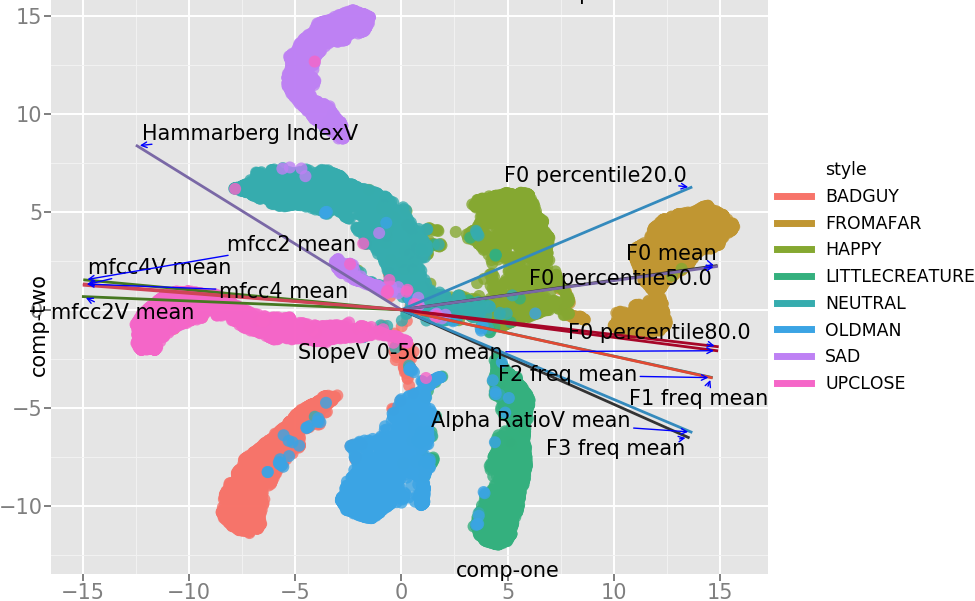}
  \caption{Latent space with directions of gradients of audio features}
  \label{fig:gradients}
\end{figure}

Dimensionality reduction is then performed with UMAP to have a collection of 2D points that one can visualize (see Figure~\ref{fig:gradients}). We observed that points in a given area of space correspond to a given style.

To have an interpretation of the space, we analyzed the trend of variation of some audio features.
To that aim, for each audio feature, i.e. the F0 mean is computed for each utterance of the dataset. Each F0 mean corresponds to one of the 2d-points of the collection.
A plane $F0_{mean} = f(x,y) =ax+by+c $ is then approximated.

We hope this plan $f(x,y)$ to be a good approximation of $F0_{mean}$. If it is, it means that going in a certain direction of the space will increase F0 mean linearly.
We check that by computing the correlation between the approximations $f(x,y)$ with the real values of F0 mean.

The gradient of the plane is 
$(\frac{\partial f}{\partial x}=a, \frac{\partial f}{\partial y}=b)$
and corresponds to the direction of its greatest slope. It thus gives the direction of the variation of the audio feature approximated by the plane.

The embedding space trained to condition a TTS system has several advantages: it does not require any label about the style and it is already in use with a TTS system.


\section{Synthesis}
\label{synthesis}

As stated in Section~\ref{related_work}, a recurrent problem while using deep learning algorithms is the need of a large amount of data, especially in fields in which data collection is costly and difficult. This is the case of emotional TTS field. In what follows, we aim to study the impact of fine-tuning applied to emotional TTS.

\subsection{Synthesis with Emotion adaptation}

In this study~\cite{exploring_transfer_learning-19-tits}, we aim to find out whether it is possible to obtain an emotional TTS system by fine-tuning a neutral TTS system with a small emotional speech dataset. We study the impact of this fine-tuning on the intelligibility of generated speech and the subjective perception of the generated speech.

The experiments were performed with an open source implementation\footnote{https://github.com/Kyubyong/dc\_tts} of DCTTS~\cite{dctts-17-tachibana}. 

A first model is trained with LJ-speech dataset, a dataset that consists of sentences uttered by a female speaker, with a total of 24 hours of speech. 

To obtain emotional TTS models by fine-tuning, we proceed in two steps. The data used is a subset of EmoV-DB corresponding to one speaker. The two steps are:
\begin{itemize}
    \item Fine-tuning of the model with the neutral part of the subset
    \item Fine-tuning copies of the resulting model separately with subsets of each emotional categories
\end{itemize}

To evaluate the first step, we use an objective measure of intelligibility proposed in~\cite{intellig-16-Orozco} on 100 sentences synthesized with three systems: the first model trained on LJ-speech dataset, the model fine-tuned with the neutral part of the subset and a model trained only with the neutral part of the subset. The average accuracies are respectively $0.630 \pm 0.042$, $0.517 \pm 0.048$, and $0.004 \pm 0.004$. These results suggest that the fine-tuning on a small subset tends to degrade a little the intelligibility. It also suggests however that using only the small dataset leads to unintelligble speech like sounds, therefore this fine-tuning is useful to generate more intelligible speech.

To evaluate the second step, we performed a subjective MOS test to measure the perception of the emotion perceived for each emotional model. The results show that the emotion are perceivable although less intelligible.



\subsection{Controlling the intensity of emotional categories}

As a demonstration, we developed a multi-emotional TTS system with the possibility to control the intensity of emotional categories. We implemented a modified version of DCTTS that takes an encoding of the emotion category at the input of the decoder. During training, a simple one-hot encoding is used. But at synthesis stage, we can modify the intensity of an emotion category by inputting encodings that are not one-hot.

The demonstration is available online\footnote{https://github.com/numediart/speak\_with\_style\_demo}.

In this demonstration, only the number corresponding to neutral and the category with which we interpolate are non-zero. The sum of the numbers of a code is one.



\section{Conclusion \& Future Works}
\label{ccl}

We propose a methodology for building a system for speech synthesis that is able to control the expressiveness of speech from user control in reall time, thanks to deep learning algorithms.

Our contribution to answer the need for data is the collection of EmoV-DB that will allow to experiment on emotional TTS.

To address the challenging task of extracting emotion/expressiveness features from speech, we present systems able to extract such a representation with supervised and unsupervised techniques. We also provide a way to visualize the trends of audio features in an embedding space.

As for synthesis itself, we studied the impact of transfer learning from neutral TTS to emotional TTS and demonstrated the benefits of this technique. 

We now plan to build a system that will take advantage of  neutral and emotional speech datasets along with the unsupervised expressiveness representation provided by our TTS algorithm, based on a variational autoencoder (VAE-TTS).


We plan to develop a multi-speaker system with VAE-TTS, to be able to share knowledge from several datasets. The challenge will be to study if the system is able to generalize expressiveness across different speaker identities, which will require it to distinguish the effects of phonetics, speaker characteristics and expressiveness on the generated speech.

We will then integrate the system in a control interface, to provide users with a friendly-user tool to generate speech with various control levels (text, speaker identity and expressiveness).

Our ultimate goal is to use this tool in a conversational context in which the system has to detect the emotional expressiveness on its interlocutor and answer him/her by imitating his/her style applied to some predefined answer.

\section*{Acknowledgment}

No\'e Tits is funded through a PhD grant from the Fonds pour la Formation \`a la Recherche dans l'Industrie et l'Agriculture (FRIA), Belgium. 

\bibliographystyle{IEEEtran}
\bibliography{IEEEabrv,refs}

\end{document}